\documentclass[nopacs,twocolumn,preprintnumbers,notitlepage,amsmath,amssymb,superscriptaddress]{revtex4-2}

\usepackage{hyperref}
\usepackage[utf8]{inputenc}
\usepackage{physics}
\usepackage{blindtext}
\usepackage{ dsfont }
\usepackage{tikz}
\usepackage{subfiles}
\usepackage{subcaption}
\usepackage[font=small,labelfont=bf,
   justification=justified,
   format=plain]{caption} 
\graphicspath{ {./Figures/} }
\usepackage{graphicx}
\usepackage{dcolumn}
\usepackage{bm}
\usepackage{amsmath}
\usepackage{amssymb}
\newcommand{\Lim}[1]{\lim\limits_{#1}}
\newcommand{\Sim}[1]{\mathrel{\mathop{\kern 0pt\sim}\limits_{#1}}}
\newcommand{\To}[1]{\mathrel{\mathop{\kern 0pt\to}\limits_{#1}}}
\newcommand{\ju}[1]{\textcolor{black}{#1}}
\makeatletter
\AtBeginDocument{\let\LS@rot\@undefined}
\makeatother
\usepackage[]{pdfpages}

\begin{document}

\title{Beyond Poisson: First-Passage Asymptotics of Renewal Shot Noise}
\author{J. Br\'emont}
\affiliation{Collège de France, 3 Rue d'Ulm, 75005 Paris, France}
\date{\today}
\begin{abstract}
The first-passage time (FPT) of a stochastic signal to a threshold is a fundamental observable across physics, biology, and finance. While renewal shot noise is a canonical model for such signals, analytical results for its FPT have remained confined to the Poisson (Markovian) case, even though non-Poisson arrival statistics are common in systems from neuronal spiking to gene expression. Here, we overcome this long-standing limitation by deriving a universal asymptotic formula for the mean FPT (MFPT) $\langle T_b \rangle$ to reach level $b$ for renewal shot noise with arbitrary arrival statistics and exponential marks. Our central result is a simple, closed-form expression that exposes the physical mechanism by which temporal correlations in arrivals modulate the baseline Arrhenius law. We show that bursty arrivals introduce universal scaling corrections that markedly accelerate threshold crossings. In turn, non-bursty arrivals remain Arrhenius-like, directly linking temporal burstiness to Arrhenius scaling. Furthermore, we \ju{show} and confirm numerically that the full FPT distribution becomes exponential at large thresholds, implying that $\langle T_b \rangle$ provides a complete asymptotic characterization. Our work, enabled by a novel exact expression for the moments of the noise, establishes a general framework for analyzing extreme events in non-Markovian systems with relaxation.
\end{abstract}
\maketitle

Threshold-crossing events driven by stochastic jump-decay processes are ubiquitous across physics, biology, and finance. In neurons, spikes occur when membrane voltage exceeds a threshold between synaptic inputs~\cite{GerstnerKistler2002,burkittReviewIntegrateandfire,drosteExactAnalytical,rijalExactDistribution,robinModelingEmergent}; in gene expression, bursty mRNA/protein levels must cross regulatory thresholds to trigger phenotypic switching~\cite{PeccoudYcart1995,FriedmanCaiXie2006PRL,ShahrezaeiSwain2008PNAS,GoldingEtAl2005Cell,RajVanOudenaarden2008Cell,meyerGeometryInducedBurstingDynamics2012,szavits-nossanSolvingStochastic}; in materials science, stress fluctuations trigger yielding events with relaxation between avalanches~\cite{zhuTemperatureStrainRate,sollichRheologySoft}; and in finance, barrier crossings determine option pricing and ruin probabilities~\cite{riceRandomNoiseI,leveilleCompoundTrend}. In all these contexts, the first-passage time (FPT) of the noise process $X(t)$ to a threshold $b$ is the central observable.

\begin{figure}
    \centering
    \includegraphics[width=\columnwidth]{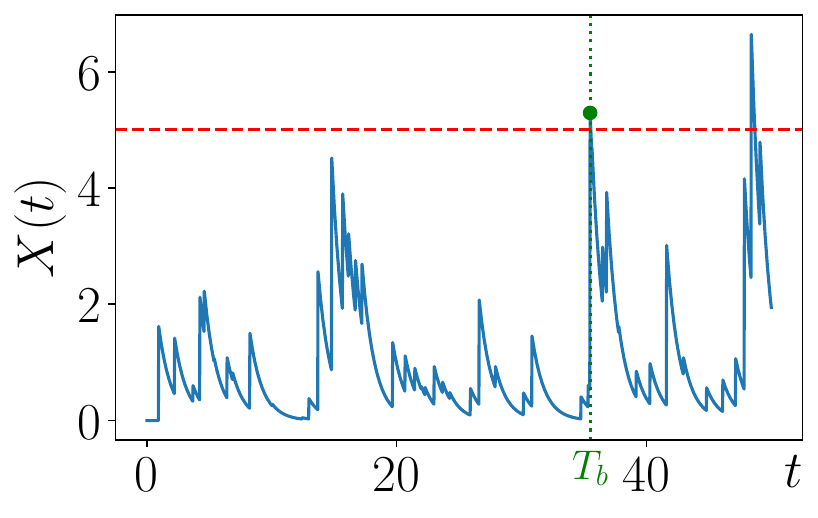}
    \caption{A typical realization of renewal shot noise $X(t)$, showing exponential relaxation between random impulsive events. The vertical green line signals the FPT $T_b$, where $X(t)$ 
exceeds the threshold $b=5$ (horizontal red line) for the first time.
    }
    \label{fig:illus}
\end{figure}

The natural model capturing these dynamics is renewal shot noise, defined by
\begin{equation*}
    X(t)=\sum_{t_i\le t} x_i\,e^{-\gamma (t-t_i)},
\end{equation*}
where $x_i$ are i.i.d. marks and interarrival times $\tau_i = t_{i+1}-t_i$ are i.i.d. with density $w(\tau)$ (see FIG.~\ref{fig:illus}). This model embodies two essential features: impulsive bursts at random times, and relaxation between events. The classical Poisson case ($w(\tau)=r e^{-r \tau}$) renders $X(t)$ Markovian, and its FPT statistics are well understood~\cite{borovkovExitTimes,kellaHittingTimes,laioMeanFirst,tsuruiFirstpassageProblem,masoliverFirstpassageTimes}. However, many applications exhibit strongly non-Poissonian arrival statistics, such as refractory periods in neuronal spiking or bursty transcription in gene expression~\cite{degerStatisticalProperties,kumarTranscriptionalBursting,nicolasDeformationFlow,olsenPartialStochastic,gallistelNeuronFunction,meyerGeometryInducedBurstingDynamics2012}, which render the process genuinely non-Markovian. Despite decades of study, analytical progress on FPT statistics has remained confined to the Poisson case, with non-Poisson shot noise presenting a long-standing challenge, as is often the case for non-Markovian processes~\cite{bremontPersistenceExponentsa,levernierSurvivalProbabilityStochastic2019,guerinMeanFirstpassage,hanggiNonMarkovProcesses}. While general, exact integral equations satisfied by the MFPT are known \cite{masoliverFirstpassageTimes}, these have proven to be intractable beyond the Poisson case: \ju{subsequent studies \cite{hernandez-garciaFirstpassageTime,dubkovProbabilityCharacteristics,bauleExponentialIncrease,weissFirstpassageTimes} did not extract closed-form asymptotic behavior from them.}

\begin{figure}
    \centering
    \includegraphics[width=\columnwidth]{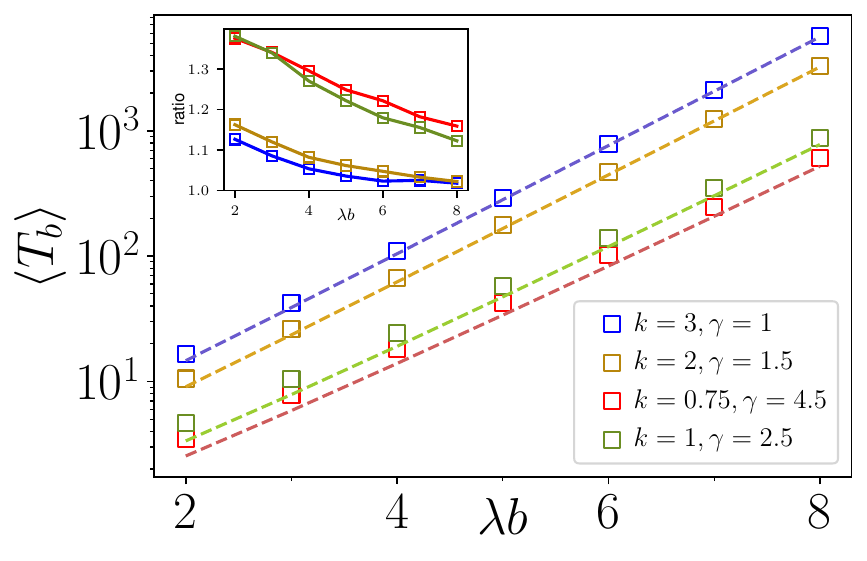}
    \caption{MFPT $\langle T_b\rangle$ of renewal shot noise with exponential marks (symbols: simulations; dashed lines: exact asymptotics, Eq.~\eqref{mfpt-general}). 
Interarrival times follow a Gamma distribution
$w(t)=\tfrac{rk}{\Gamma(k)}(rk t)^{k-1}e^{-rk t}$ with \ju{shape parameter $k$ and mean rate chosen as} $r=1.2/k$, while marks are exponential with $\lambda=1$.
\ju{The blue and yellow cases} highlight refractory effects ($w(0)=0$), \ju{whereas the green curve and the red Poisson baseline} illustrate bursty dynamics \ju{($w(0)>0$ or infinite)}.
Statistical errors are smaller than the symbol size.
\ju{The inset shows the ratio $\langle T_b\rangle/\langle T_b\rangle_{\mathrm{RE}}$,
where $\langle T_b\rangle_{\mathrm{RE}}=1/(r p(b))$ is the rare-event estimate introduced in Eq.~\eqref{mfpt-asymp} and given by Eq.~\eqref{mfpt-general}.
In agreement with the analysis presented in the SM, one observes $\langle T_b\rangle\ge \langle T_b\rangle_{\mathrm{RE}}$ at finite $b$.
As shown in the inset, the convergence $\langle T_b\rangle/\langle T_b\rangle_{\mathrm{RE}}\to1$ is slower in the bursty regime $\kappa\le1$, consistent with enhanced clustering of the shots.}
}
    \label{fig:mfpt}
\end{figure}

\begin{figure}
    \centering
    \includegraphics[width=\columnwidth]{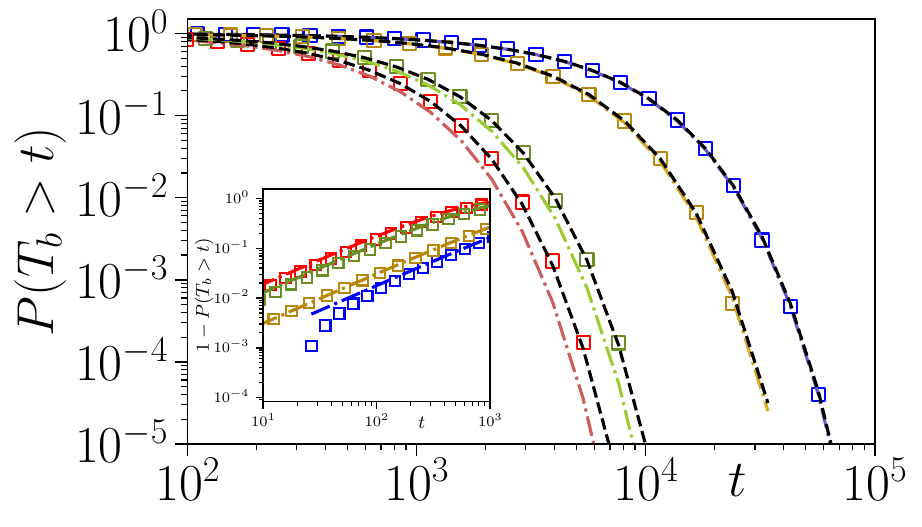}
    \caption{Numerical confirmation of the exponential form of the FPT distribution at large thresholds.
Shown is the cumulative distribution function $\mathbb{P}(T_b>t)$ obtained from simulations (symbols) for $b=8$ and $\lambda =1$.
\ju{The color code corresponds to different Gamma distributions for the interarrival times and is the same as in FIG.~\ref{fig:mfpt}.}
Dotted lines show the exponential distribution \eqref{exp-dist-tb}:
black lines use the empirical mean $\langle T_b\rangle$ measured in simulations, while colored lines use the theoretical MFPT predicted by Eq.~\eqref{mfpt-general}.
The excellent agreement, in particular between the black lines and the simulation data, provides strong numerical confirmation of our result \eqref{exp-dist-tb} that the distribution of $T_b$ becomes exponential for large $b$.
\ju{The inset displays $1-\mathbb{P}(T_b>t)$ in the small-$t$ regime, confirming the exponential form already well before $T_b$ approaches its mean value $\langle T_b\rangle$, which, as discussed in the SM, emerges rapidly. In contrast, the convergence of $\langle T_b\rangle$ to the rare-event estimate Eq.~\eqref{mfpt-general} is slower in the bursty regime $\kappa\le1$.}
}
    \label{fig:fpt-dist-exp}
\end{figure}

In this Letter, we derive, to our knowledge, the first exact asymptotic expression for the MFPT $\langle T_b \rangle$ valid for general renewal arrivals and exponential marks. Our result, given by a compact product formula Eq. \eqref{mfpt-general}, reveals explicitly how general interarrival statistics impact the FPT scaling, and in particular reveals universal deviations from the baseline Arrhenius law. This breakthrough is enabled by a novel closed-form expression for the Laplace transform of all moments of $X(t)$, a result of independent interest that provides a powerful analytical tool for studying renewal shot noise. \\
\textit{Main result.} We now present our main result for the MFPT $\langle T_b \rangle$. The marks are taken to be exponentially distributed, $x_i \sim \mathrm{Exp}(\lambda^{-1})$, and the arrival process has a finite mean rate $r \equiv \left(\int_0^\infty t\, w(t)\,dt\right)^{-1}$. The process may start at any value $0 \leq X(0) \ll b$.
For a function $f(t)$, we denote its Laplace transform by $\hat f(s)=\int_0^\infty e^{-st} f(t)\,dt$.
Our central result is the following simple, exact asymptotic expression for the MFPT at large thresholds \footnote{We write $A(x) \sim B(x)$ for $x\to \infty$ if the ratio $A(x)/B(x) \to 1$. The product term $\prod_{m=1}^{\lambda b}\!\bigl[1-\hat{w}(m\gamma)\bigr]$ is defined via analytic continuation when $\lambda b \notin \mathbb{N}$. See SM for more details and an explicit analytical continuation in the case of Gamma distributed interarrivals considered on FIGs~\ref{fig:mfpt} and \ref{fig:fpt-dist-exp}.}:
\begin{equation}
\label{mfpt-general}
\langle T_b\rangle \;\sim\;
\frac{\exp(\lambda b)}
{r} \prod_{m=1}^{\lambda b}\!\bigl[1-\hat{w}(m\gamma)\bigr],
\qquad b\to\infty.
\end{equation}
This key formula, confirmed numerically in FIG.~\ref{fig:mfpt}, leads to several important insights.
(i) We check that Eq.~\eqref{mfpt-general} reduces to known results in important limits. First, for Poisson arrivals, $\hat{w}(s)=r/(s+r)$, it simplifies to the classical asymptotic expression~\cite{borovkovExitTimes,kellaHittingTimes,laioMeanFirst,tsuruiFirstpassageProblem,masoliverFirstpassageTimes}:
\begin{equation}
\label{mfpt-poisson}
\langle T_b^{\mathrm{Poisson}}\rangle \underset{b\to\infty}{\sim}
\frac{1}{\gamma}\,\Gamma\!\left(\tfrac{r}{\gamma}\right)\,
(\lambda b)^{-r/\gamma}\,e^{\lambda b}.
\end{equation}
Second, in the limit of instantaneous relaxation ($\gamma \to \infty$), we have $\hat{w}(m\gamma) \to 0$, yielding the pure Arrhenius law $r \langle T_b \rangle \sim e^{\lambda b} = 1/\mathbb{P}(x > b)$. Indeed, in this limit, each impulse is an upcrossing with probability $\mathbb{P}(x > b) = e^{-\lambda b}$.
(ii) Equation~\eqref{mfpt-general} extends MFPT asymptotics for shot noise far beyond the Poisson case, providing (to our knowledge) the first closed analytic form valid for general renewal arrivals and exponential marks. Comparable asymptotics are extremely rare for non-Markovian processes~\cite{hanggiNonMarkovProcesses,guerinMeanFirstpassage,bremontPersistenceExponentsa,brayPersistenceFirstPassage}.
(iii) Although the full distribution of $T_b$ for arbitrary $b$ remains an open problem for non-Markovian shot noise, our results provide a complete asymptotic characterization in the large-threshold limit.
\ju{Indeed, we show in the Supplementary Material (SM) \cite{suppmatbib} that for $b\to\infty$, the FPT $T_b$ becomes exponentially distributed:}
\begin{equation}
    \label{exp-dist-tb}
    \mathbb{P}(T_b > t)\;\underset{b \to \infty}{\sim}\; \exp\!\left(-t / \langle T_b \rangle\right),
\end{equation}
where $\langle T_b \rangle$ is now fully explicit \eqref{mfpt-general}. \ju{Physically, this reflects the fact that upcrossings of a high threshold become asymptotically independent events.
This property is well established for Poisson shot noise~\cite{borovkovExitTimes,kellaHittingTimes} and more generally arises in rare-event limits~\cite{glassermanLimitsFirst,chupeauCoverTimesRandom2015}.}
\ju{Our numerical simulations in FIG.~\ref{fig:fpt-dist-exp} provide clear confirmation of this asymptotic behavior.} 
Thus, our result \eqref{mfpt-general} fully quantifies FPT statistics of shot noise with exponential marks in the large-threshold limit.
(iv) As detailed in the \emph{Physical interpretation} section, Eq.~\eqref{mfpt-general} exposes a universal mechanism governing the kinetics of rare threshold crossings. The MFPT retains an Arrhenius-like scale $\langle T_b \rangle \propto e^{\lambda b}$, but the product term in Eq.~(1) quantifies how temporal correlations (here, in the arrival process) modulate this baseline exponential scale, which is a central question in first-passage kinetics \cite{hanggiNonMarkovProcesses,barbier-chebbahLongtermMemory,wallaceNonArrheniusKineticsLoop2001}. Assume the interarrival density behaves for short times as $w(t)\sim c\,t^{\kappa-1}$ with $\kappa>1/2$ (for $0<\kappa\leq 1/2$, Arrhenius modulation persists but leads to more intricate corrections detailed in the SM). Our expression \eqref{mfpt-general} then yields the universal asymptotic scaling
\begin{equation}
\label{mfpt-explicit}
\langle T_b \rangle \underset{b \to \infty}{\sim} \frac{B \, e^{\lambda b}}{r}\times
\begin{cases}
e^{-\tfrac{c\,\Gamma(\kappa)}{(1-\kappa)\,\gamma^{\kappa}}\left(\lambda b\right)^{1-\kappa}}, & \tfrac{1}{2}<\kappa<1,\\[4pt]
(\lambda b)^{-c/\gamma}, & \kappa=1,\\[4pt]
1, & \kappa>1,
\end{cases}
\end{equation}
where the prefactor $B=B[w(t),\gamma]$ (given in the SM) is nonuniversal and depends on the full interarrival law. The scaling itself, however, is universal and dictated solely by the short-time behavior of $w(t)$ through the parameters $c$ and $\kappa$. Physically, when $w(0)=0$ ($\kappa>1$), short interarrival times are suppressed, akin to refractory periods in neuronal spiking. In this regime, exponential relaxation prevents the accumulation of impulses, and threshold crossings are dominated by a single, rare large mark of size $\mathcal{O}(b)$, resulting in pure Arrhenius scaling. Conversely, when short gaps are prevalent ($\kappa \leq 1$, with the baseline Poisson case included in $\kappa=1$), as observed in e.g. bursty gene transcription \cite{meyerGeometryInducedBurstingDynamics2012,szavits-nossanSolvingStochastic}, crossings are driven by bursts of arrivals rather than isolated events. This cooperative mechanism leads to substantial acceleration of threshold crossings: algebraic for $\kappa = 1$, and stretched-exponential for $\kappa < 1$, thereby establishing a direct quantitative link between arrival burstiness and suppression of the Arrhenius timescale $e^{\lambda b}/r$. As shown in the SM, this correction strengthens as $\kappa$ decreases.
In the representative case $\kappa = 1/2$, relevant to transport-limited gene dynamics~\cite{meyerGeometryInducedBurstingDynamics2012} through the return statistics of a one-dimensional random walk, the interarrival density behaves as $w(t) \sim ct^{-1/2} + d$.
The MFPT then acquires an additional polynomial correction: $\langle T_b \rangle \sim B e^{\lambda b - 2c\sqrt{\tfrac{\pi\lambda b}{\gamma} }} (\lambda b)^{-d/\gamma}$.

\textit{Derivation of \eqref{mfpt-general}.}
The derivation of our main result \eqref{mfpt-general} is based on a new, exact formula for the moments of $X(t)$, a result of independent interest. Despite extensive study of shot-noise \cite{blanterButtikerShotNoise,boxmaShotnoiseQueueing,jangMomentsRenewal,leveilleCompoundTrend}, closed-form results for the moments of renewal shot noise remain limited. The Poisson case, where $X(t)$ is Markovian, is classical and well understood \cite{coxPointProcesses,grandellAspectsRisk}, but for general arrivals, analysis has typically stopped at the first two moments \cite{coxSuperpositionRenewal,grandellAspectsRisk,degerStatisticalProperties}. Higher moments have previously been addressed through recursive schemes in actuarial mathematics \cite{LeveilleGarrido2001,jangMomentsRenewal}, with much of this line of analysis dating back to Takács \cite{takacsStochasticProcesses}, but such schemes become unwieldy at increasing order. In neuroscience, where shot noise corresponds to postsynaptic currents in the so-called Stein model \cite{Stein1965,burkittReviewIntegrateandfire,GerstnerKistler2002} and non-Poisson arrival statistics are well established \cite{degerStatisticalProperties}, analysis typically relies on Gaussian or diffusion approximations \cite{burkittReviewIntegrateandfire} which are known to fail outside narrow parameter regimes \cite{burkittReviewIntegrateandfire,degerStatisticalProperties}. Although more detailed pool-based synaptic release and network models have been analyzed in detail \cite{brunelFastGlobal,rijalExactDistribution}, the renewal shot noise statistics treated here are more general and not contained in those models. Consequently, despite decades of use across disciplines, no compact closed-form expression for higher moments has been available.

We now present a closed-form expression for the Laplace transform of $\langle X(t)^n \rangle$, valid for arbitrary renewal processes and mark distributions with finite moments. Let \(\mu_k \equiv \langle x_i^k \rangle\) and $\hat{\psi}(s) \equiv \hat{w}(s)/[1-\hat{w}(s)]$. We show in SM the following exact expression:
\begin{align}
    \label{moments-general}
     \widehat{\langle X(t)^n \rangle} = \frac{1}{s+n\gamma} &\sum_{\substack{n_1 + \dots + n_k = n \\ 1\leq k \leq n}} \binom{n}{n_1, \dots, n_k} \times \nonumber \\
     &\prod_{m=1}^k \mu_{n_m} \cdot \hat{\psi}\left(s+\gamma\sum_{j=1}^{m-1} n_j \right).
\end{align}
For the crucial case of exponential marks ($\mu_m = m!\,\lambda^{-m}$), this general result collapses to a remarkably simple product form:
\begin{equation}
\label{moments-exp}
\widehat{\langle X(t)^n \rangle}
= \frac{\hat{\psi}(s)\,\lambda^{-n}\,n!}{s+n\gamma}
    \prod_{m=1}^{n-1} \left(1 + \hat{\psi}(s+m\gamma)\right).
\end{equation}


To the best of our knowledge, neither \eqref{moments-general} nor its specialization \eqref{moments-exp} has appeared in the literature. We recognize the latter as closely connected to the factorial moments of the $G/M/\infty$ queue \cite{takacsStochasticProcesses,szavits-nossanSolvingStochastic} (see SM), but it does not appear in recent reviews of shot noise \cite{boxmaShotnoiseQueueing}. These expressions offer a broadly applicable, fully analytical alternative to the recursive or approximate methods commonly used. In particular, the final-value theorem $\Lim{s \to 0} s \hat{f}(s) = \Lim{t \to \infty} f(t)$ applied to \eqref{moments-general} gives exact moments of all orders in the stationary state $X(t \to \infty) \equiv X_\infty$ of the shot noise. As shown below, it is precisely these stationary moments that will allow us to compute the MFPT. Our starting point is the following rare-event estimate for the MFPT to threshold $b$:
\ju{
\begin{equation}
    \label{mfpt-asymp}
    \langle T_b \rangle \Sim{b \to \infty} \langle T_b \rangle_{\mathrm{RE}} \equiv \frac{1}{r\,p(b)} ,
\end{equation}
}
where $p(b)$ is the probability that a single impulse, sampled from the stationary pre-jump distribution, pushes the process above the threshold $b$.
\ju{The convergence of the actual MFPT $\langle T_b \rangle$ to this rare-event estimate is demonstrated in the SM, where it is shown that the generic inequality $\langle T_b \rangle \geq \langle T_b \rangle_{\mathrm{RE}}$ holds at finite $b$ (see FIGs.~\ref{fig:mfpt} and \ref{fig:fpt-dist-exp}). We provide a physical interpretation of Eq.~\eqref{mfpt-asymp} here: it rests on three key points.}
(i) For large $b$, threshold crossings are rare and occur on timescales much larger than the time needed for the process to reach stationarity (which is independent of $b$). Thus, crossings take place in the stationary regime.
(ii) In this rare-event limit, multiple crossings occurring within a short time window are exponentially unlikely, implying that an upcrossing of $b$ at time $t$ is overwhelmingly likely to be the first one.
(iii) In the stationary regime, all impulses are statistically equivalent, so that the mean number of threshold crossings per unit time is simply $r\,p(b)$.
\ju{Although this rare-event estimate is formally reminiscent of Kac-type relations \cite{kacNotionRecurrence,benichouFirstpassageTimes}, which relate return times to a state $x$ to the inverse of the stationary density $1/P_s(x)$, we stress that Eq.~\eqref{mfpt-asymp} is of a different nature, as it concerns first-passage events rather than returns to the threshold $b$.}
The quantity of interest is now
\begin{equation}
    \label{pofb-def}
    p(b) \equiv \mathbb{P}(X^+_\infty > b,\;X^-_\infty<b),
\end{equation}
where $X^-_\infty$ (resp. $X^+_\infty$) denotes the shot noise just before (resp. after) an impulse in the stationary regime. Importantly, except in the case of Poisson arrivals, $X^-_\infty$ is not distributed as $X_\infty$. Because the mark $X^+_\infty-X^-_\infty$ is exponentially distributed with mean $1/\lambda$, Eq.~\eqref{pofb-def} can be written as
\begin{equation}
    p(b) = e^{-\lambda b}\,\int_0^b e^{\lambda x}\,P(X^-_\infty=x)\,dx,
\end{equation}
where we introduced the truncated moment–generating function (mgf) of the pre-burst shot noise.  
In the SM we show that a computation analogous to Eq.~\eqref{moments-general} gives the stationary pre-burst moments
\begin{equation}
    \label{moments-stat-xm}
    \langle (X^-_\infty)^n \rangle
    = \hat\psi(n\gamma)\,\lambda^{-n}\,n!\,\prod_{m=1}^{n-1}\!\left(1+\hat \psi(m\gamma)\right), 
    \qquad n \geq 1.
\end{equation}
The final step relies on an asymptotic duality (derived in SM) between truncated moment sums and integrals: for a \ju{nonnegative} random variable $Y$ with mgf $\langle e^{t Y} \rangle$ finite for $t<\lambda$,
\begin{equation}
    \label{duality}
    \sum_{n=0}^{\lambda b} \frac{\langle (\lambda Y)^n \rangle}{n!}
    \;\underset{b \gg 1/\lambda}{\sim}\; 
    \int_0^b e^{\lambda y} P(Y=y)\,dy.
\end{equation}
Applying \eqref{duality} to $Y = X_\infty^-$ and noting the telescopic identity for $n \geq 1$,
\begin{equation}
    \frac{\langle (\lambda X^-_\infty)^n \rangle}{n!} = B_{n+1} - B_n, 
    \quad B_n \equiv \prod_{m=1}^{n-1} [1-\hat{w}(m\gamma)]^{-1},
\end{equation}
the sum \ju{in the LHS of} \eqref{duality} collapses, yielding Eq.~\eqref{mfpt-general}.

\textit{Physical interpretation.}
We emphasize that the product form in Eq.~\eqref{mfpt-general} is not a mere mathematical artifact but reflects a clear physical mechanism, presented below. 
For large thresholds $b$, the MFPT is given by \ju{the rare-event estimate $\langle T_b\rangle \sim \langle T_b\rangle_{\mathrm{RE}}= 1/(r\,p(b))$}. To understand the product structure physically, we analyze how the crossing probability changes when we increment the threshold by one mark unit: $b^+ \equiv b+1/\lambda$. The ratio $p(b^+)/p(b)$ represents the conditional probability to cross $b^+$ given that $b$ has been crossed, and is governed by two distinct scenarios.
Consider the overshoot $x_0$ remaining immediately after crossing $b$. Crucially, for exponential marks, $x_0$ is also exponentially distributed due to the memoryless property. This leads to two possible mechanisms:
(S1) With probability $e^{-1}$, we have $x_0 > 1/\lambda$, meaning the same impulse that crossed $b$ also suffices to cross $b^+$. This contributes a term $e^{-1}p(b)$ and is responsible for the Arrhenius factor $e^{\lambda b}$.
(S2) If instead $x_0 \leq 1/\lambda$, the process starts below $b^+$ after crossing $b$. Since $X(t)$ rarely sits near such high values, crossing $b^+$ typically occurs through a burst of additional impulses before significant relaxation below $b$ can occur. This burst mechanism explains deviations from pure Arrhenius scaling.
More precisely, in scenario (S2), crossing $b^+$ is achieved by $n \geq 1$ additional impulses with amplitudes $x_1,\dots,x_n$ arriving at times $t_1,\dots,t_n$. The interarrival times $\tau_i = t_i-t_{i-1}$ are most probably smaller than $(\lambda b^+ \gamma)^{-1}$, the time needed for $X(t)$ starting from $b^+$ to relax by one mark unit $1/\lambda$. Since $\gamma\tau_i \ll 1$ for these relevant timescales, the condition for crossing $b^+$ after exactly $n$ additional impulses becomes:
\begin{equation}
A_n = 
\begin{cases}
0 \leq \sum_{k=0}^i x_k - \gamma b^+ \sum_{k=1}^i \tau_k < \tfrac{1}{\lambda}, & 0\leq i<n, \\
\sum_{k=0}^n x_k - \gamma b^+ \sum_{k=1}^n \tau_k \geq \tfrac{1}{\lambda}. &
\end{cases}
\end{equation}
This condition ensures that the process remains between $b$ and $b^+$ until the final impulse pushes it above $b^+$.
The total weight of scenario (S2) is therefore:
\begin{multline}
    \label{scenario-general-mfpt}
\sum_{n=1}^\infty \int_0^\infty dt \int_{\tau_1+\cdots+\tau_n=t} 
w(\tau_1)\cdots w(\tau_n)\,\mathbb{P}(A_n).
\end{multline}
While this expression holds for general mark distributions, it simplifies dramatically for exponential marks. In this case,
\begin{equation}
\mathbb{P}(A_n) = e^{-\lambda\gamma b^+ \left[\sum_{k=1}^n \tau_k\right] - 1}.
\end{equation}
This allows the $n$-fold integral in \eqref{scenario-general-mfpt} to decouple into a product of individual Laplace transforms, yielding a geometric sum of terms $e^{-1}\hat{w}(\lambda b^+ \gamma)^n$ in \eqref{scenario-general-mfpt}. Combining both scenarios yields the recursion relation:
\begin{equation}
\label{heuristic-exp}
\frac{p(b^+)}{p(b)}\underset{b\to\infty}{\sim}\frac{1}{e}\left(1+\frac{\hat{w}(\lambda b^+ \gamma)}{1-\hat{w}(\lambda b^+ \gamma)}\right),
\end{equation}
which is exactly equivalent to our main result in Eq.~\eqref{mfpt-general}. This physical picture enables two key insights. First, it clarifies why extending the explicit MFPT beyond the exponential-mark case solved here appears out of reach: for general marks, the overshoot law is unknown and $\mathbb{P}(A_n)$ lacks a closed form. Second, because only interarrival times with $\gamma\tau_i \ll 1$ contribute in the large-$b$ limit, our exact result Eq.~\eqref{mfpt-general} depends solely on the local decay rate following each impulse, not on the full exponential form $x_i e^{-\gamma\tau_i}$. Consequently, Eq.~\eqref{mfpt-general} is robust to the precise form of the relaxation mechanism: it remains valid if the exponential relaxation is replaced by any smooth relaxation with the same initial slope $-\gamma x_i$ (see \cite{burenevFirstpassageProperties} for the constant-slope case).

\color{black}
\textit{Conclusion.}
We have characterized the first-passage time $T_b$ to a high threshold $b$ for non-Markovian renewal shot noise with exponential marks.
Our main results are: (i) an explicit asymptotic formula for the MFPT $\langle T_b\rangle$ (Eq.~\eqref{mfpt-general}), proven exact as $b \to \infty$ in the SM; (ii) the full FPT distribution becomes exponential for large $b$, i.e. $P(T_b > t) \sim e^{-t/\langle T_b \rangle}$ (Eq.~\eqref{exp-dist-tb}), confirmed analytically (see SM) and numerically (FIG.~\ref{fig:fpt-dist-exp}). By providing an explicit expression for the MFPT $\langle T_b\rangle$, our results fully characterize the distribution of $T_b$; (iii) exact closed-form expressions for all moments of renewal shot noise (Eqs.~\eqref{moments-general}–\eqref{moments-exp}), previously unavailable in the literature.
These results expose how temporal correlations, that is, burstiness or refractoriness in arrivals, universally modulate the baseline Arrhenius scaling (Eq.~\eqref{mfpt-explicit}), establishing a direct link between microscopic arrival statistics and macroscopic extreme-event kinetics.
Our framework enables quantitative predictions of rare threshold crossings in diverse non-Markovian systems, from neuronal spiking to gene expression bursts.
A natural next step is quantifying finite-threshold corrections beyond the rare-event estimate $\langle T_b\rangle_{\mathrm{RE}}$, particularly in the bursty regime where such corrections could be of importance in quantifying phenotypic switching in transport-limited gene dynamics~\cite{meyerGeometryInducedBurstingDynamics2012,szavits-nossanSolvingStochastic}.

\color{black}

\ju{
Details and source code for the numerical simulations used to generate
FIGs.~\ref{fig:mfpt} and \ref{fig:fpt-dist-exp} are available at
\href{https://github.com/julien-bremont/Shot-noise}{\color{blue}{https://github.com/julien-bremont/Shot-noise}}.
}


%



\section*{Supplementary material}

# Supplementary Material : Beyond Poisson: First-Passage Asymptotics of Renewal Shot Noise

J. Brémont[1]

[1] Collège de France, 3 Rue d'Ulm, 75005 Paris, France

(Dated: October 24, 2025)

## I. DERIVATION OF THE MAIN MOMENTS FORMULA

In this section we prove Eq. (1) of the main text. We begin with an important lemma.

### A. A lemma about multi-time integrals in Laplace space

Say we want to compute the following integral

$$I(t) \equiv \int_0^t dt_1 \int_{t_1}^t dt_2 \cdots \int_{t_{n-1}}^t dt_n f_1(t_1) f_2(t_2 - t_1) \ldots f_n(t_n - t_{n-1}) e^{\gamma(t_1 + \cdots + t_n)}, \tag{1}$$

where $\gamma > 0$ and $f_i$ are some functions of time. This expression can be more naturally expressed as the integral of a convolution over the variables $u_i \equiv t_i - t_{i-1}$, where we defined $t_0 = 0$:

$$I(t) = \int_0^t d\sigma \int_0^\sigma f_1(u_1) \ldots f_n(u_n) e^{\gamma(nu_1 + (n-1)u_2 + \cdots + u_n)} \delta(u_1 + \cdots + u_n - \sigma) du_1 \ldots du_n. \tag{2}$$

It is now straightforward to consider the Laplace-transformed version

$$\hat{I}(s) = \int_0^\infty e^{-st} I(t) dt. \tag{3}$$

Using standard Laplace transform manipulations, we see from (2) that

$$\hat{I}(s) = \frac{\prod_{i=1}^n \hat{f}_i(s - (n+1-i)\gamma)}{s}. \tag{4}$$

### B. Proof of Eq. (1)

We make use of the representation

$$X(t) = \sum_{m=0}^\infty \mathbf{1}_{\{\tau_m < t\}} e^{-\gamma(t - \tau_m)} b_m, \tag{5}$$

where the arrival times $\{\tau_n\}$ are assumed to be increasingly sorted. Equivalently, we write

$$e^{\gamma t} X(t) = \sum_{m=0}^\infty \mathbf{1}_{\{\tau_m < t\}} e^{\gamma \tau_m} b_m. \tag{6}$$

Let us write the $k$-th power of $e^{\gamma t} X(t)$ as

$$e^{n\gamma t} X(t)^n = \sum_{m_1, \ldots, m_n = 0}^\infty \mathbf{1}_{\{\tau_{m_1} < t, \ldots, \tau_{m_n} < t\}} e^{\gamma(\tau_{m_1} + \cdots + \tau_{m_n})} b_{m_1} \ldots b_{m_n}. \tag{7}$$

First, we take the mean over the random amplitudes $b_i$ and regroup factors $b_{m_i}$ that have the same index $m_i$. We obtain

$$e^{n\gamma t} \mathbb{E}_{\{b_i\}}[X(t)^n] = \sum_{m_1, \ldots, m_n = 0}^\infty \mathbf{1}_{\{\tau_{m_1} < t, \ldots, \tau_{m_n} < t\}} e^{\gamma(k_1 \tau_{m_1} + \cdots + k_n \tau_{m_n})} \langle b^{n_1} \rangle \ldots \langle b^{n_n} \rangle, \tag{8}$$

where $n_i$ is the number of occurences of the integer $m_i$ in the $n$-uple $(m_1, \ldots, m_n)$. Because of the permutation symmetry of the summand, it is now natural to sum over strictly increasingly sorted $k$-uples $(m_1 < \cdots < m_k)$, with $1 \leq k \leq n$, which introduces a multinomial factor:

$$e^{n\gamma t} \mathbb{E}_{\{b_i\}}[X(t)^n] = \sum_{k=1}^{n} \sum_{0 \leq m_1 < \cdots < m_k < \infty} \sum_{\substack{n_1 + \cdots + n_k = n \\ 1 \leq n_i \leq n}} \binom{n}{n_1, \ldots, n_k} \mathbf{1}_{\{\tau_{m_1} < \cdots < \tau_{m_k} < t\}} e^{\gamma(n_1 \tau_{m_1} + \cdots + n_k \tau_{m_k})} \langle b^{n_1} \rangle \ldots \langle b^{n_k} \rangle. \tag{9}$$

It now remains to average the above over arrival times $\tau_i$ and Laplace transform $t \to s$. We thus need to compute the following Laplace transform

$$\int_0^\infty e^{-st} \langle \sum_{0 \leq m_1 < \cdots < m_k < \infty} \mathbf{1}_{\{\tau_1 < \cdots < \tau_k < t\}} e^{\gamma(n_1 \tau_1 + \cdots + n_k \tau_k)} \rangle_{\{\tau_i\}} dt. \tag{10}$$

First, we represent the sum above as an integral over each time of arrival :

$$\sum_{0 \leq m_1 < \cdots < m_k < \infty} \mathbf{1}_{\{\tau_1 < \cdots < \tau_k < t\}} e^{\gamma(n_1 \tau_1 + \cdots + n_k \tau_k)} = \int_0^t dt_1 \int_{t_1}^t dt_2 \cdots \int_{t_{k-1}}^t dt_k \delta(\tau_1 - t_1) \ldots \delta(\tau_k - t_k) e^{\gamma(n_1 t_1 + \cdots + n_k t_k)}.$$

Taking the expectation of the above with respect to the arrival times $\tau_i$ and using our lemma (4), we can write (10) as

$$\frac{1}{s} \prod_{i=1}^{k} \hat{\psi}\left(s - \gamma \sum_{j=1}^{i} n_k\right), \tag{11}$$

where $\hat{\psi}(s) \equiv \hat{w}(s)/(1 - \hat{w}(s))$ is the Laplace transform of the renewal density. Finally, taking care of the final $e^{-n\gamma t}$ term coming from (9), which shifts the Laplace variable $s \to s + n\gamma$, we obtain our main result, Eq. (1) of the main text :

$$\widehat{\langle X(t)^n \rangle} = \frac{1}{s + n\gamma} \sum_{\substack{n_1 + \cdots + n_k = n \\ 1 \leq k, n_i \leq n}} \binom{n}{n_1, \ldots, n_k} \times \prod_{m=1}^{k} \mu_{n_m} \cdot \hat{\psi}\left(s + \gamma \sum_{j=1}^{m-1} n_j\right).$$

For exponentially distributed marks $\langle b^n \rangle = n! \lambda^n$, we see that the weight $\sum_{\substack{n_1 + \cdots + n_k = n \\ 1 \leq k, n_i \leq n}} \binom{n}{n_1, \ldots, n_k} \times \prod_{m=1}^{k} \mu_{n_m}$ of each summand in (12) does not depend on the explicit partition $n_1 + \cdots + n_k$. We can thus rewrite

$$\widehat{\langle X(t)^n \rangle} = \frac{\hat{\psi}(s) \lambda^n n!}{s + n\gamma} \prod_{m=1}^{n-1} \left(1 + \hat{\psi}(s + m\gamma)\right). \tag{12}$$

Indeed, in the product term of (12), all partitions of $n$ appear exactly once : partition $(n_1, \ldots, n_k)$ is obtained by choosing the term $\hat{\psi}(s + m\gamma)$ for $m \in \{n_1, \ldots, n_k\}$, and choosing the term 1 for other integers. Furthermore, it is clear that the exponential distribution of marks is the only distribution which yields such a compact product term by weighing each partition the same.

### C. Stationary moments

Isolating the term $\hat{\psi}(s) \underset{s \to 0}{\sim} \frac{r}{s}$ where $r$ is the mean interarrival rate, we apply the final value theorem $\lim_{s \to 0} s\hat{f}(s) = \lim_{t \to \infty} f(t)$ to (12). This yields the exact stationary moments

$$\langle X(t)^n \rangle \underset{t \to \infty}{\to} \langle X_\infty^n \rangle = \frac{r}{n\gamma} \sum_{\substack{n_1 + \cdots + n_k = n \\ 1 \leq k, n_i \leq n}} \binom{n}{n_1, \ldots, n_k} \left(\prod_{m=1}^{k} \mu_{n_m}\right) \prod_{m=2}^{k} \hat{\psi}\left(\gamma \sum_{j=1}^{m-1} n_j\right). \tag{13}$$

For exponentially distributed marks, we obtain

$$\langle X(t)^n \rangle \underset{t \to \infty}{\to} \langle X_\infty^n \rangle = \frac{r \lambda^n (n-1)!}{\gamma} \prod_{m=1}^{n-1} \left(1 + \hat{\psi}(m\gamma)\right). \tag{14}$$

Note that for Poisson arrivals, we recover the Gamma distribution as a limiting distribution.

## II. PROOF OF THE ASYMPTOTIC DUALITY BETWEEN ORDER-TRUNCATED AND VARIABLE-TRUNCATED MGFS

We want to show that for a nonnegative random variable $Y$ with asymptotically monotonous smooth density $P(y)$ and finite moment generating function $\langle e^{tY}\rangle$ for $t < \lambda$,

$$\sum_{n=0}^{\lambda b} \frac{\langle (\lambda Y)^n\rangle}{n!} \underset{b \gg 1/\lambda}{\sim} \int_0^b e^{\lambda y} P(y)\, dy. \tag{15}$$

### Step 1. Trivial case

If $\langle e^{\lambda Y}\rangle < \infty$, then both sides converge to this finite limit as $b \to \infty$, and the statement is immediate. The interesting regime is $\langle e^{\lambda Y}\rangle = +\infty$.

### Step 2. Rewrite the sum as an integral with a kernel

Exchanging sum and integral (all terms positive),

$$\sum_{n=0}^{\lambda b} \frac{\langle (\lambda Y)^n\rangle}{n!} = \int_0^\infty K_b(y)\, e^{\lambda y} P(y)\, dy, \tag{16}$$

where

$$K_b(y) = e^{-\lambda y} \sum_{n=0}^{\lambda b} \frac{(\lambda y)^n}{n!} = \Pr\bigl(\mathrm{Poisson}(\lambda y) \le \lambda b\bigr). \tag{17}$$

Thus the sum is the Laplace integral, but with a smoothed cutoff $K_b(y)$ instead of the sharp cutoff $\mathbf{1}_{\{y \le b\}}$.

### Step 3. Behavior of the kernel

The kernel $K_b(y)$ is the cumulative distribution of a Poisson variable:

- For $y < b$, the mean $\lambda y$ is below the cutoff, so $K_b(y) \approx 1$.
- For $y > b$, the mean is above the cutoff, so $K_b(y) \approx 0$.
- The transition from 1 to 0 occurs only in a narrow window of width $\sim \sqrt{b}$ around $y = b$. Outside this window, large-deviation estimates for the Poisson distribution give

$$K_b(y) \asymp \exp[-\lambda y\, I(b/y)], \qquad I(x) = x\log x - x + 1 > 0,$$

  where $I$ is the Poisson rate function, so the kernel is exponentially close to either 0 (for $y < b$) or 1 (for $y > b$.).

In short, $K_b(y)$ acts like a smoothed step function at $y = b$.

### Step 4. Negligible contributions

Decompose

$$\int_0^\infty K_b(y) e^{\lambda y} P(y)\, dy = \underbrace{\int_0^b e^{\lambda y} P(y)\, dy}_{C} + \underbrace{\int_b^\infty K_b(y) e^{\lambda y} P(y)\, dy}_{A} - \underbrace{\int_0^b (1 - K_b(y)) e^{\lambda y} P(y)\, dy}_{B}. \tag{18}$$

For $y > b$, $K_b(y)$ is exponentially small, so $A \ll C$. For $y < b$, $1 - K_b(y)$ is exponentially small, so $B \ll C$. The boundary layer $|y - b| \lesssim \sqrt{b}$ contributes only a negligible fraction compared with the bulk growth of $C$, due to the asymptotic monotonous behavior of $P(y)$.

**Step 5. Conclusion**

Since $C = \int_0^b e^{\lambda y} P(y)\, dy \to \infty$ as $b \to \infty$ (because the full integral diverges at $\lambda$), both correction terms are negligible. Therefore

$$\sum_{n=0}^{\lambda b} \frac{\langle (\lambda Y)^n \rangle}{n!} \sim \int_0^b e^{\lambda y} P(y)\, dy, \tag{19}$$

which proves Eq. (15).

**Why does $X_\infty^-$ verify the hypotheses ?**

To show that the mgf $\langle e^{t X_\infty^-} \rangle$ is finite for $t < \lambda$, given the moments of $X_\infty^-$ shown in the main text, it suffices to show that for $t < \lambda$ one has

$$\sum_{n=0}^{\infty} (t/\lambda)^n \hat{\psi}(n\gamma) \prod_{m=1}^{n-1} (1 + \hat{\psi}(m\gamma)) < \infty. \tag{20}$$

Writing $\prod_{m=1}^{n-1}(1 + \hat{\psi}(m\gamma)) = e^{\sum_{m=1}^{n-1} \log(1+\hat{\psi}(m\gamma))}$, we see that (20) holds because $\sum_{m=1}^{n} \hat{\psi}(m\gamma) \ll n$.

### III. COMPUTING THE MFPT IN THE CASE OF EXPONENTIAL MARKS

#### A. Per-arrival success $p(b)$

We now specialize to exponential marks $b_i \sim \text{Exp}(\lambda^{-1})$. Our starting point is the rare-event estimate for the MFPT to threshold $b$:

$$\langle T_b \rangle \underset{b \to \infty}{\sim} \frac{1}{r\, p(b)}, \tag{21}$$

where $r$ is the mean interarrival rate, defined by $\hat{\psi}(s) \underset{s \to 0}{\sim} r/s$, and $p(b)$ is the probability that a single impulse in the stationary state pushes the process above the threshold $b$. Equation (21) rests on three key points:

1. For very large $b$, crossings are very rare, so the process is near stationarity when a crossing occurs.

2. Multiple crossings are exponentially less likely than a single crossing in the large-$b$ limit, so a crossing of $b$ at some time $t$ is very likely to be the first one.

3. In the stationary regime, all impulses are equivalent, so the mean number of crossings per unit time is $r\, p(b)$.

We write

$$p(b) \equiv \mathbb{P}(X_\infty^+ > b,\ X_\infty^- < b) = \int_0^b \mathbb{P}(X_\infty^+ - X_\infty^- \geq b - x^-)\, f_{X^-}(x^-)\, dx^-, \tag{22}$$

where $X_\infty^-$ (resp. $X_\infty^+$) denotes the shot noise just before (resp. after) an impulse in the stationary regime, and $f_{X^-}$ its density. For non-Poisson arrivals, $X_\infty^-$ is *not* distributed as the unconditional stationary $X_\infty$, so we need its moments explicitly. A calculation analogous to Eq. (12), with conditioning on an arrival at the final observation time, gives the moments of the pre-burst shot noise $X_\infty^-$ as

$$\langle (X_\infty^-)^n \rangle = \hat{\psi}(n\gamma)\, \lambda^n\, n! \prod_{m=1}^{n-1} (1 + \hat{\psi}(m\gamma)). \tag{23}$$

We check that in the case of Poisson arrivals $\hat{\psi}(s) = \frac{r}{s}$ (and only in this case), the statistics of the variable $X_\infty^-$ conditioned right before a mark, given by (23), are exactly that of the unconditioned variable $X_\infty$ given by (14).

Because the marks are exponential, the increment $X_\infty^+ - X_\infty^-$ is exponential as well. Thus

$$p(b) = \int_0^b e^{-(b-x)/\lambda}\, f_{X^-}(x)\, dx = e^{-b/\lambda}\, \mathbb{E}\!\left[e^{X_\infty^-/\lambda};\ X_\infty^- < b\right]. \tag{24}$$

## B. The prefactor $B$

Assume the interarrival density behaves for short times as $w(t) \sim c\,t^{\kappa-1}$ with $\kappa > 1/2$. Then, the Laplace transform behaves as $\hat{w}(s) \underset{s\to\infty}{\sim} \Gamma(\kappa)s^{-\kappa}$. Then, we easily deduce from our main result Eq. (1) of the main text that the prefactor $B$ introduced by Eq. (5) of the main text reads

$$B = \begin{cases} \prod_{m=1}^{\infty}[1 - \hat{w}(m\gamma)], & \kappa > 1 \\ \lim_{n\to\infty}\left[n^{c/\gamma}\prod_{m=1}^{n}[1 - \hat{w}(m\gamma)]\right], & \kappa = 1 \\ \lim_{n\to\infty}\left[e^{\frac{c\,\Gamma(\kappa)}{(1-\kappa)\gamma^{\kappa}}n^{1-\kappa}}\prod_{m=1}^{n}[1 - \hat{w}(m\gamma)]\right], & \frac{1}{2} < \kappa < 1. \end{cases} \qquad (25)$$

Indeed, let us do the case $\frac{1}{2} < \kappa < 1$ as an example. We have

$$\langle T_b \rangle \sim \frac{\exp(\lambda b)}{r}\prod_{m=1}^{\lambda b}[1 - \hat{w}(m\gamma)]. \qquad (26)$$

The large-$m$ expansion of $\hat{w}(m\gamma)$ reads

$$\hat{w}(m\gamma) = c\Gamma(\kappa)(m\gamma)^{-\kappa} + O(m^{-1-\varepsilon}),\ \varepsilon = 2\kappa - 1 > 0. \qquad (27)$$

Hence, the product term in (26) becomes, for large $\lambda b$

$$\prod_{m=1}^{\lambda b}[1 - \hat{w}(m\gamma)] = \exp\left(A - c\sum_{m=1}^{\lambda b}\Gamma(\kappa)(m\gamma)^{-\kappa}\right)e^{O((\lambda b)^{-\varepsilon})} \underset{\lambda b \to \infty}{\sim} \exp\left(A - c\sum_{m=1}^{\lambda b}\Gamma(\kappa)(m\gamma)^{-\kappa}\right). \qquad (28)$$

Because the sum in the exponential is not convergent as $\lambda b \to \infty$ for $\kappa < 1$, it can be approximated by its continuous, integral version up to a constant $C$ (from e.g. the Euler-Maclaurin formula)

$$\sum_{m=1}^{\lambda b}\Gamma(\kappa)(m\gamma)^{-\kappa} = C + \int_{1}^{\lambda b}\Gamma(\kappa)(m\gamma)^{-\kappa}dm + o(1). \qquad (29)$$

This yields exactly (25) for the case $\frac{1}{2} < \kappa < 1$, using the correct constant $B$.

The case $\kappa \leq 1/2$ is harder to treat as it introduces corrections which depend on higher orders of the expansion of $w(t)$ close to $t = 0$. Let us do the $\kappa = 1/2$ case as an example, by choosing, as $t \to 0$,

$$w(t) = ct^{-1/2} + d + O(t^{1/2}). \qquad (30)$$

One has

$$\hat{w}(m\gamma) = \frac{\sqrt{\pi}c}{\sqrt{m\gamma}} + \frac{d}{m\gamma} + O(m^{-3/2}). \qquad (31)$$

Hence, the product term in (26) becomes

$$\prod_{m=1}^{\lambda b}[1 - \hat{w}(m\gamma)] = \exp\left(A - \sum_{m=1}^{\lambda b}\left[c\sqrt{\pi}(m\gamma)^{-1/2} + \frac{d}{m\gamma}\right]\right)e^{O((\lambda b)^{-1/2})} \underset{\lambda b \to \infty}{\sim} \exp\left(C - 2c\sqrt{\frac{\pi\lambda b}{\gamma}} - \frac{d}{\gamma}\log(\lambda b)\right). \qquad (32)$$

Hence, the MFPT behaves as $\langle T_b \rangle \asymp e^{\lambda b - 2c\sqrt{\frac{\pi\lambda b}{\gamma}}}(\lambda b)^{-d/\gamma}$, with an additional polynomial correction from the $\kappa > 1/2$ case.

## IV. SHOT NOISE VS. $G^X/M/\infty$: LINK, WHAT IS KNOWN, AND WHY THE APPROACHES DIFFER

The $G^X/M/\infty$ is defined as such [1]. Let arrivals form a renewal process with arrival times $\{t_n\}$ (i.e, the interarrival times $t_{n+1} - t_n$ are i.i.d with density $w$), with random i.i.d batch sizes $B_n$ of customers arriving at time $t_n$, and exponential service rate $\gamma > 0$; each customer $i$ has a *random* service time $S_i \sim \text{Exp}(\gamma)$.

**Link between the shot noise $X$ and the queue length $N$.** Given $\{(\tau_n, B_n)\}$, customers in batch $n$ are served after time $t$ independently with probability $p_n = e^{-\gamma(t-t_n)}$. Thus

$$N(t) \mid \{(\tau_n, B_n)\} = \sum_n \text{Binomial}(B_n, p_n), \qquad \mathbb{E}_{\{S_i\}}\big[N(t) \mid \{(\tau_n, B_n)\}\big] = X(t), \tag{33}$$

where we used the mean value $B_n p_n$ of the binomial distribution $\text{Binomial}(B_n, p_n)$. So, if the impulse amplitude at time $\tau_n$ is $B_n$, $X(t)$ is exactly the service-time average of $N(t)$. However, at the distributional level the two are unrelated: knowing the pmf $\mathbb{P}\{N(t) = k\}$ is *not* enough to reconstruct the pdf of $X(t)$. In general one would need the full conditional law $N(t) \mid \{S_i\}$, averaged over all service times. This "de-Poissonization" problem is typical in probability theory. This is why the stationary laws of $N$ and $X$ have very different levels of tractability: while the stationary pmf of $N$ dates back to [2] in the single-batch case $B_n \equiv 1$, which is the queue model noted $G/M/\infty$, no closed stationary pdf of $X$ exists in general.

**Stationary results: $N$ (Takács) vs. $X$.** For the $G/M/\infty$ queue (unit batches, $B_n \equiv 1$), Takács showed that the stationary queue length $N(\infty)$ admits an explicit closed-form discrete distribution [2]:

$$P\big(N(\infty) = m\big) = \frac{r}{\gamma} \sum_{j=m}^{\infty} (-1)^{j-m} \binom{j}{m} \frac{1}{j} \prod_{i=1}^{j-1} \hat{\psi}(i\gamma), \qquad \hat{\psi}(s) = \frac{\hat{w}(s)}{1 - \hat{w}(s)}, \quad r = \frac{1}{\int_0^\infty t\, w(t)\, dt}, \quad m \geq 1. \tag{34}$$

This expression follows from the fact that the probability generating function

$$U(z) = \sum_{m=0}^{\infty} P(m)\, z^m$$

satisfies a solvable fixed-point equation (see [1, 2] for the derivation). The derivatives $U^{(n)}(1)$ give the factorial moments of the stationary queue length $N(\infty)$, from which $P(N(\infty) = m)$ can be determined explicitly. These factorial moments are [2]

$$\langle N(\infty)_k \rangle = \langle N(\infty)(N(\infty) - 1) \cdots (N(\infty) - k + 1) \rangle = r\, \frac{(k-1)!}{\gamma} \prod_{i=1}^{k-1} \hat{\psi}(i\gamma). \tag{35}$$

Interestingly, these factorial moments are closely related to our stationary-moment formula (14) for the exponential shot noise with unit-mean marks, through the formal replacement $\hat{\psi} \to 1 + \hat{\psi}$. The origin of this correspondence is not fully understood, but since $X$ and $N$ are linked by (33), such a relation is perhaps not too surprising.

---



\end{document}